\documentclass[amssymb,prb,twocolumn,floats,amsmath,showpacs,superscriptaddress]{revtex4}
\usepackage{bm}
\usepackage{graphicx}

\begin{document}

\title{Excitonic giant Zeeman effect in GaN:Mn$^{3+}$}
%
%
\author{W.~Pacuski}
\email{Wojciech.Pacuski@fuw.edu.pl}

\affiliation{Institut N\'eel/CNRS-Universit\'e J. Fourier, Bo\^{\i}te Postale 166, F-38042 Grenoble Cedex 9, France}

\affiliation{Institute of Experimental Physics, Warsaw University,
Ho\.za 69, PL-00-681 Warszawa, Poland}

\author{D.~Ferrand}\email{david.ferrand@grenoble.cnrs.fr}
\affiliation{Institut N\'eel/CNRS-Universit\'e J. Fourier, Bo\^{\i}te Postale 166, F-38042 Grenoble Cedex 9, France}

\author{J.~Cibert}
\affiliation{Institut N\'eel/CNRS-Universit\'e J. Fourier, Bo\^{\i}te Postale 166, F-38042 Grenoble Cedex 9, France}

\author{J.~A.~Gaj}

\affiliation{Institute of Experimental Physics, Warsaw University,
Ho\.za 69, PL-00-681 Warszawa, Poland}

\author{A.~Golnik}
\affiliation{Institute of Experimental Physics, Warsaw University,
Ho\.za 69, PL-00-681 Warszawa, Poland}

\author{P.~Kossacki}
\affiliation{Institute of Experimental Physics, Warsaw University,
Ho\.za 69, PL-00-681 Warszawa, Poland}

\author{S.~Marcet}
\affiliation{Institut N\'eel/CNRS-Universit\'e J. Fourier, Bo\^{\i}te Postale 166, F-38042 Grenoble Cedex 9, France}

\author{E.~Sarigiannidou}\altaffiliation[Permanent address: ]{LMGP BP 257- INPGrenoble Minatec - 3 parvis Louis N\'eel - 38016 Grenoble} \affiliation{Institut N\'eel/CNRS-Universit\'e J. Fourier, Bo\^{\i}te Postale 166, F-38042 Grenoble
Cedex 9, France}

\author{H.~Mariette}
\affiliation{Institut N\'eel/CNRS-Universit\'e J. Fourier, Bo\^{\i}te Postale 166, F-38042 Grenoble Cedex 9, France}
%
%
%
\begin{abstract}
We describe a direct observation of the excitonic giant Zeeman
splitting in Ga$_{1\textrm{-}x}$Mn$_{x}$N, a wide-gap III-V diluted
magnetic semiconductor.
Reflectivity and absorption spectra measured at low temperatures
display the $A$ and $B$ excitons, with a shift under magnetic field
due to $s$,$p$-$d$ exchange interactions. Using an excitonic model,
we determine the difference of exchange integrals between Mn$^{3+}$
and free carriers in GaN, ${N_0(\alpha-\beta)}=-1.2\pm0.2$~eV.
Assuming a reasonable value of $\alpha$, this implies a positive
sign of $\beta$ which corresponds to a rarely observed ferromagnetic
interaction between the magnetic ions and the holes.
\end{abstract}
%
%
%
%
\maketitle
%
%
Diluted magnetic semiconductors (DMS) of the III-V type, such as (Ga,Mn)As, attract attention mainly because of their magnetic and electron transport
properties. The direct measurement of the strength of the ion-carrier coupling, as routinely performed in II-VI DMS, through the observation of the
giant Zeeman splitting of excitons, was not accessible for III-V DMS due to free carriers introduced by magnetic ions. Hence conclusions have been
based on studies of extremely diluted samples \cite{dilGaAs} or on a complex interpretation of experimental data.\cite{complexGaAs,Okab98,Hwan05}

Here we show and quantitatively describe excitons in GaN, with a
shift due to the $s$,$p$-$d$ coupling to Mn ions incorporated as
neutral centers (Mn$^{3+}$). We present both the magneto-optical
study of specially designed and thoroughly characterized
Ga$_{1-x}$Mn$_{x}$N layers, and the excitonic model needed to
determine the strength of ion-carrier coupling. Such a model is
needed, because excitonic effects are especially strong in wide-gap
semiconductors. Hence the giant Zeeman effect measured on the
excitons cannot be identified with that of band-to-band transitions,
as it has been done for other wurtzite DMS.\cite{Agga83}


Ga$_{1\textrm{-}x}$Mn$_{x}$N layers of about 400~nm-thick were grown by molecular beam epitaxy on sapphire substrates with a GaN or AlN buffer layer,
along the $c$-axis perpendicular to the surface \cite{Kuro03}. Details on the growth can be found in Ref. 7. Layers grown with a GaN buffer allowed
to study quantitatively\cite{Marc06b} the infrared absorption due Mn$^{3+}$, as identified in quasi-bulk crystals,\cite{Wolo04b} and they exhibit
\cite{Marc06b} magnetic circular dichroism (MCD) at the bandgap energy,\cite{Ando03} due to the ion-carrier coupling. Layers with a higher Mn content
evidence a ferromagnetic behavior at low temperature.\cite{Sari06} The present Ga$_{1\textrm{-}x}$Mn$_{x}$N layers have been grown on a buffer layer
made of AlN, which has a large energy gap (over 6~eV), so that the excitonic spectra of Ga$_{1\textrm{-}x}$Mn$_{x}$N ($E_g$ about 3.5~eV) are not
entangled with those of the buffer layer. The presence of the AlN buffer layer with a lattice mismatch introduces strain and disorder, which
increases the excitonic linewith. As disorder increases also with the Mn content, we focused our study onto rather dilute samples ($x<1.2\%$)
exhibiting well-resolved excitons.


The concentration of Mn$^{3+}$ has been determined using near
infrared spectroscopy. We observe sharp absorption lines previously
reported\cite{Wolo04b,Marc06b} as intra-ionic $^5T_2\rightarrow
{^5E}$ transitions of Mn$^{3+}$. The intensity of these lines has
been calibrated\cite{Marc06b} against secondary ion mass
spectrometry (SIMS) in samples, where the $d^4$ configuration of Mn
has been confirmed by x-ray absorption at the Mn
K-edge.\cite{Titov05} In the present series of samples, the
intensity of the Mn$^{3+}$ absorption follows the total Mn
concentration determined by SIMS with an accuracy better than 10\%.
Furthermore, a Mn$^{2+}$ concentration lower than 0.01\% was deduced
from the absence of the characteristic lines in electron
paramagnetic resonance spectra.\cite{Pascal06b}
%
%
\begin{figure}
\includegraphics*[width=80mm]{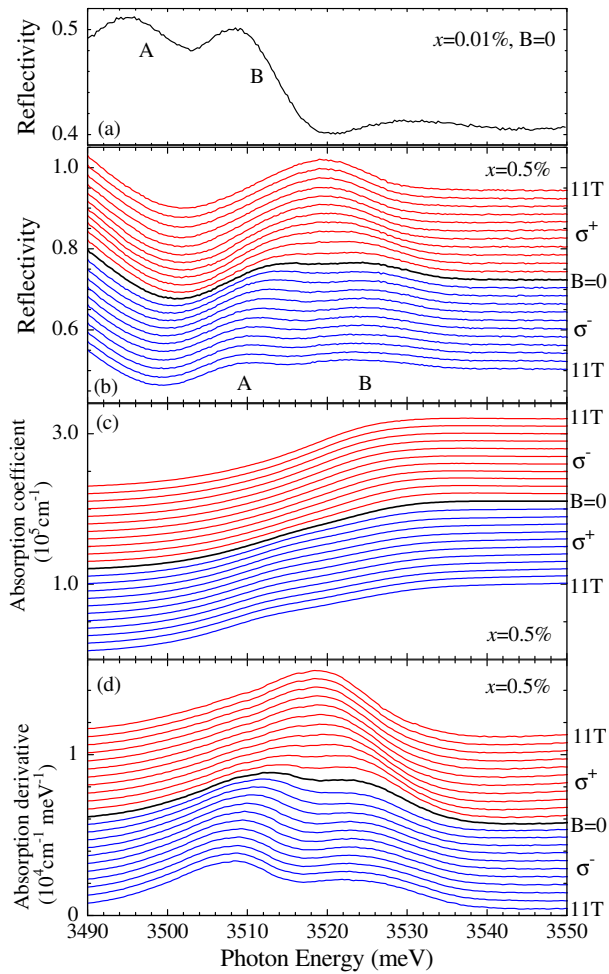}
\caption[]{(Color online) (a) Reflectivity of  $A$ and $B$ excitons
measured for Ga$_{1\textrm{-}x}$Mn$_{x}$N   with $x=0.01\%$. (b)
Reflectivity, (c) absorption and (d) derivative of absorption, for
$x=0.5\%$, in Faraday configuration, in $\sigma^+$ (top, red color)
and $\sigma^-$ (bottom, blue color) circular polarizations, at
$T=1.7$~K.} \label{fig:spectra}
\end{figure}
\begin{figure}
\includegraphics*[width=80mm]{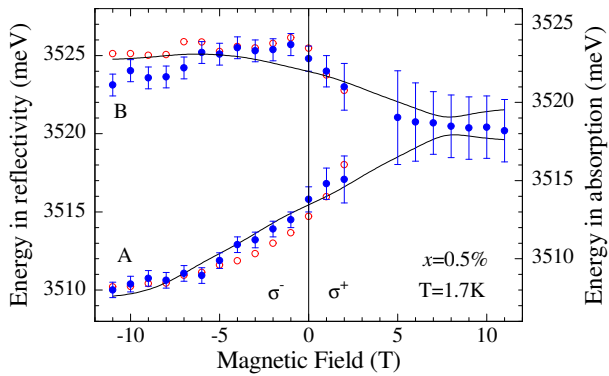}
\caption[]{(Color online) Energies of excitonic transitions
determined from reflectivity peaks (filled circles, blue color) and
from derivative of absorption (empty circles, red color), at
$T=1.7$~K and $B$$||$$c$. Corresponding spectra are shown in Figs.
\ref{fig:spectra}(b) and \ref{fig:spectra}(d). Solid lines,
excitonic shifts calculated along the model and with parameters as
discussed in text.} \label{fig:position}
\end{figure}


In wurtzite GaN, the crystal field and the spin-orbit coupling split the valence band into three components, from which three excitons are formed,
labeled $A$, $B$, and $C$ in the order of increasing energy. There is a strong optical anisotropy. If the electric field $E$ of the incident light is
perpendicular to the $c$-axis ($\sigma$ polarization), the oscillator strength is larger for excitons $A$ and $B$ than for exciton $C$. Exciton $C$
has the largest oscillator strength if $E$$\parallel$$c$ ($\pi$ polarization). The position and oscillator strength of all excitons, and particularly
exciton $C$, are modified by strain.\cite{Gil95}


We measured reflectivity at temperatures down to 1.7~K, with the
light from a high-pressure Xe lamp incident along the $c$-axis. In
this configuration only the $\sigma$ polarization is accessible. In
zero-field spectra,  we observe two characteristic features. For the
most diluted sample [${x=0.01\%}$, Fig.~\ref{fig:spectra}(a)], the
energy of these features corresponds to the energy of the $A$ and
$B$ excitons in GaN with a biaxial compressive strain of about
19~kbar using data from Ref.~\onlinecite{Gil95}. Such a strain
strongly reduces the oscillator strength of exciton $C$, which is
also particularly sensitive to strain disorder. This explains why we
do not observe a third structure at 3540~meV, as expected for
exciton $C$ at this strain value. In another sample with $x=0.5$\%
[see the middle spectra marked as $B=0$ in
Fig.~\ref{fig:spectra}(b)], the $A$ and $B$ excitonic features are
further broadened and shifted to higher energies. This shift
confirms the increase of the bandgap energy upon increasing the Mn
content, previously inferred from MCD spectra.\cite{Marc06b}


Magneto-optical spectra were measured in the Faraday configuration,
with the magnetic field parallel to the common $c$-axis and optical
axis. Magneto-reflectivity spectra of pure GaN or very diluted
GaN:Mn (${x=0.01\%}$), not shown, exhibit a very small excitonic
Zeeman shift, ${\Delta E}<0.5$~meV, even at $B$=11~T, in agreement
with previously reported data on GaN.\cite{Step99,Camp97} The Zeeman
shift of excitons is strongly enhanced in the presence of a
significant content of Mn ions. This is shown in
Fig.~\ref{fig:spectra}(b), which displays reflectivity spectra of
the sample with $x=0.5$\%, at 1.7~K and different values of the
applied field. The $\sigma^+$ and $\sigma^-$ circular polarizations
are defined with respect to the direction of the applied field.
Exciton $A$ shifts to low energy (redshift) in $\sigma^-$
polarization (${\Delta E}=6$~meV at $B$=11~T). In $\sigma^+$, it
shifts to high energy and merges with the $B$ exciton. The behavior
of exciton $B$ is more complex, as shown by the plot of the
positions of the reflectivity maxima shown in
Fig.~\ref{fig:position} (filled circles). Note also that the
saturation of the redshift of $A$ starts at quite high field
values, when compared to other DMS.


Excitons $A$ and $B$ are observed also in absorption, Fig.~\ref{fig:spectra}(c). Excitonic features are more visible in the derivative of these
absorption spectra, Fig.~\ref{fig:spectra}(d): the positions of the peaks are plotted in Fig.~\ref{fig:position} as empty circles, using the right
axis. They are shifted by 2~meV with respect to the maxima in reflectivity. The exact energy of the excitonic transitions is neither the energy of
the reflectivity maxima (see the position of transverse polaritons in Ref.~\onlinecite{Lago81}) nor that of the peaks in the derivative of the
absorption spectra. However, since the shape of the excitonic features appears to be not affected by the applied field, selecting arbitrary a
position within the observed feature results only in a constant shift. Hence we obtain a good estimate of the Zeeman shifts even if we do not know
the absolute position of the exciton. This is confirmed by the agreement between the exciton evolution determined from the two kinds of spectra
in Fig.~\ref{fig:position}.

We now turn to a detailed analysis of shifts such as observed in
Fig.~\ref{fig:position}. It includes the proper description of the
valence band and of excitons in GaN, the strength of the $s$,$p$-$d$
exchange interactions, and the magnetic properties of Mn in GaN.
%


We first briefly remind the different components entering the
excitonic Hamiltonian and discuss necessary assumptions. In
zero-field, the energy of a hole in the valence band of a
semiconductor with the wurtzite structure is given by:\cite{Juli98}
\begin{equation}
{
H_v=-\tilde\Delta_1(L_z^2\textrm{-}1)-2\Delta_2 L_z s_z-2\Delta_3(L_x s_x\textrm{+}L_y s_y),
}
\label{eq:Hv}
\end{equation}
where $\tilde\Delta_1$ describes the effect of the trigonal
components of crystal field and strain, and $\Delta_2$ and
$\Delta_3$ the anisotropic spin-orbit interaction; $L_\alpha$ and
$s_\alpha$ are projections of the orbital and  spin momenta,
respectively. The $z$ direction is parallel to the $c$-axis.

Electron-hole interactions within the exciton results in
${H_{e-h}=-R^*+2\gamma~\vec{\bold s}_e \vec{\bold s}}$,
where $R^*$ is the binding energy, $\vec{\bold s}_e$ is the electron
spin and $\vec{\bold s}$ the hole spin, and the electron-hole
exchange\cite{Juli98,Step99} is parameterized by the exchange
integral $\gamma$. A possible difference in the binding energies of
the three excitons cannot be distinguished from a change in the
values of $\tilde\Delta_1$ or $\Delta_2$ above.

Both holes and electrons interact with the Mn ions through the
so-called $s$,$p$-$d$ exchange. In this work we use the standard
form of the $s$,$p$-$d$ Hamiltonian, which only depends on the
relative orientation between the carrier spin and the averaged Mn
spin $\langle \vec{\bold S}\rangle$. It reads:
${H_{s,p\textrm{-}d}=N_0(\alpha \vec{\bold s}_e +\beta \vec{\bold
s})~x\langle -\vec{\bold S}\rangle}$,
where $\alpha$ and $\beta$ are the $s$,$p$-$d$ exchange integrals for electrons and holes, respectively. A more complex description of the $p$-$d$
interaction for the $d^4$ electronic configuration has been given in Ref. \onlinecite{Bhat94}, but our experimental precision does not allow us to
evidence any effect of the additional terms involving the orbital momentum. The direct influence of the magnetic field on the exciton (usual Zeeman
effect and diamagnetic shift) can be also easily implemented in our model,\cite{Step99} but it is small enough to be safely neglected.


Finally, our Hamiltonian takes the following form: $H=E_0 + H_v +
H_{e-h}+H_{sp-d}$,  where $E_0$ is the bandgap energy.
For a magnetic field parallel to the $c$-axis and light incident
along the same direction, we limit ourselves to $\Gamma_5$ excitons
optically active in $\sigma$ polarization. The Hamiltonian separates
into two operators acting in two subspaces corresponding
respectively to excitons active in the $\sigma^+$ and $\sigma^-$
circular polarizations. For $\sigma^+$ we use the following basis:
${|s\downarrow~p^{+}\uparrow\rangle}$ and
${|s\uparrow~p^{+}\downarrow\rangle}$ (which are identically active
in $\sigma^{+}$ polarization and will give the main contribution to
excitons $A$ and $B$), and
${|s\uparrow~p^{z}\uparrow\rangle}$ (which is optically inactive
since it is spin-forbidden, but will give the main contribution to
exciton $C$ in $\sigma$ polarization). In this basis the
Hamiltonians are written:\cite{Juli98,Pacu06a,Pacu06c}
\begin{equation}
H_v=
\left(%
\begin{array}{ccc}
  -\Delta_2 & 0         & 0 \\
  0         & \Delta_2  & -\sqrt{2}\Delta_3 \\
  0         & -\sqrt{2}\Delta_3 & \tilde{\Delta}_1\\
\end{array}%
\right),
\end{equation}
\begin{equation}
H_{e-h}=-R^*+\frac \gamma 2
\left(%
\begin{array}{ccc}
  -1    & 2     & 0 \\
  2     & -1    & 0 \\
  0     & 0     & 1 \\
\end{array}%
\right), \\
\end{equation}
\begin{equation}
H_{sp-d}^{\sigma\pm}=\pm\frac 1 2 N_0~x\langle -S_z\rangle
\left(%
\begin{array}{ccc}
  \beta-\alpha & 0 & 0 \\
  0 & \alpha-\beta& 0 \\
  0 & 0& \alpha+\beta\\
\end{array}%
\right) .\label{eq:GZ}
\end{equation}
The same matrices, with opposite giant Zeeman terms, apply in
$\sigma^{-}$ polarization with the basis
${|s\uparrow~p^{-}\downarrow\rangle}$,
${|s\downarrow~p^{-}\uparrow\rangle}$, and
${|s\downarrow~p^{z}\downarrow\rangle}$. We do not give the
corresponding matrices for the other 6 excitons ($\Gamma_1$,
$\Gamma_2$, $\Gamma_6$), which are not optically active in $\sigma$
polarizations.

\begin{figure}
\includegraphics*[width=80mm]{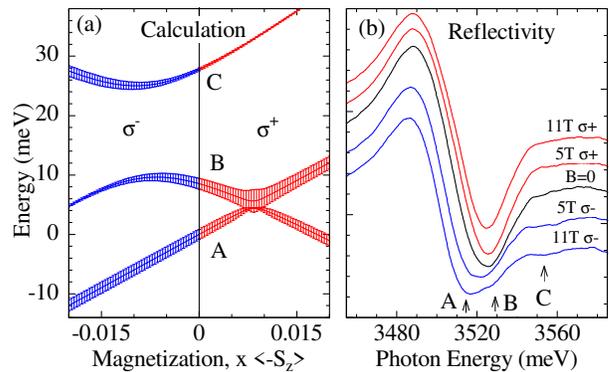}
\caption[]{(Color online)  (a) Calculated energy of the $A$, $B$,
and $C$ excitons, \textit{vs.} magnetization expressed as $x\langle
-S_z\rangle$, for $\sigma^-$ (left half) and $\sigma^+$ (right half)
circular polarizations . Oscillator strengths are shown by bars. (b)
Reflectivity of the sample with the highest Mn concentration
($x$=1.2\%). At zero field and in $\sigma^+$ polarisation excitonic
structures are no resolved, but in $\sigma^-$ polarization 3
structures can be distinguish in agreement with the model. }
\label{fig:model}
\end{figure}

Diagonalizing the Hamiltonian gives the energy of the three
excitons $A$, $B$ and $C$ in each circular polarization, as it is shown
in Fig.~\ref{fig:model}(a). The corresponding oscillator strength marked
by thickness of the line is deduced from the projection of
the corresponding eigenvector, $|\psi\rangle$, onto the relevant
subspace active in circular polarization: it is proportional to
${|\langle s\downarrow p^{\pm}\uparrow|\psi\rangle + \langle
s\uparrow p^{\pm}\downarrow |\psi \rangle|^2}$, for $\sigma^\pm$
circular polarization, respectively.


Parameters used in Fig.~\ref{fig:model}(a) were determined as follows:
using ${\Delta_2 = 6.2}$~meV, ${\Delta_3 = 5.5}$~meV and
$\gamma$=0.6~meV as reported for pure GaN,\cite{Gil95,Juli98} and
keeping $\tilde\Delta_1$ and $N_0(\alpha-\beta)$ as free parameters,
we fit the data in Fig.~\ref{fig:position} using the mean spin of
Mn$^{3+}$ given in Ref~\onlinecite{Marc06b}. We found
${\tilde\Delta_1=17}$~meV, larger than in relaxed GaN
($\Delta_1=10$~meV) due to compressive strain,\cite{Gil95} and
${N_0(\alpha-\beta)=-1.2\pm0.2}$~eV. To plot Fig.~\ref{fig:model}(a)
we used ${N_0\alpha}=0.2$~eV, a typical value in DMS.


In zero field, the relative position of the excitons is determined
mainly by the trigonal component of the crystal field including
strain, $\tilde{\Delta}_1$, and the parallel spin-orbit interaction
$\Delta_2$. The giant Zeeman shift of the exciton is induced by
$s$,$p$-$d$ interactions. The
electron-hole exchange interaction $\gamma$ governs the anticrossing
and mixing of the $A$ and $B$ excitons. It strongly alters the
excitonic oscillator strength. The perpendicular spin orbit interaction
($\Delta_3$) is responsible for the anticrossing and mixing between
the $B$ and $C$ excitons. As a result, exciton $C$ acquires a strong
oscillator strength if the giant Zeeman splitting is large enough,
as in the sample in Fig.~\ref{fig:model}(b).


\begin{figure}
\includegraphics*[width=80mm]{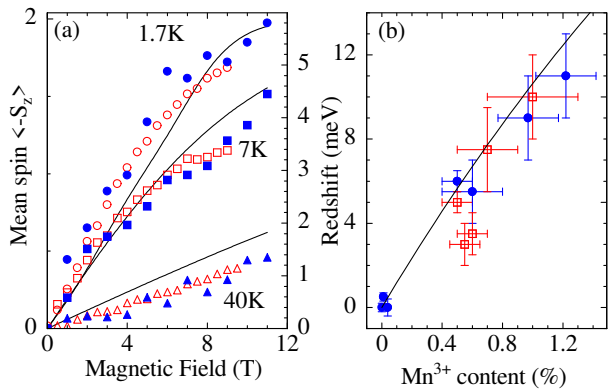}
\caption[]{(Color online) (a) Redshift of the $A$ exciton determined
using reflectivity (full symbols, blue color) and absorption (empty
symbols, red color), compared to the mean spin of Mn$^{3+}$ (solid
lines, right axis) calculated using Ref.~\onlinecite{Marc06b}, for
temperatures 1.7, 7, and 40 K. (b) Giant Zeeman shift of exciton $A$
at saturation, determined using reflectivity (full circles, blue
color) and absorption (empty squares, red color). Solid line
represents value of the shift ${\Delta
E_A=-1/2N_0(\alpha-\beta)x_\textrm{eff}\langle -S_z\rangle}$
calculated with $N_0(\alpha-\beta)$=-1.2~eV, $\langle
-S_z\rangle$=2, $x_\textrm{eff}$=$x(1-x)^{12}$. } \label{fig:shift}
\end{figure}
%
An important result is that the giant Zeeman splitting of $A$ and
$B$ excitons is not proportional to the Mn magnetization. In order
to probe the giant Zeeman energy [$N_0(\alpha-\beta)~x\langle
-S_z\rangle$], one should rather use the redshift of exciton $A$, which
appears to be proportional to the magnetization as shown by the
straight line in Fig.~\ref{fig:model}(a).

The redshift of $A$ is well accounted for a large range of fields
and temperatures, as shown in Fig.~\ref{fig:shift}(a), where the
mean spin of Mn$^{3+}$ is calculated using
Ref.~\onlinecite{Marc06b}. Both saturate only at the lowest
temperature and highest magnetic field, as the field is applied
along the hard magnetization axis of the strongly anisotropic
Mn$^{3+}$ ion. Any contribution of Mn$^{2+}$ ions would saturate at
low field (3~T), as does the Brillouin function for $d^5$. We do not
observe such a contribution, in agreement with our estimate of a
negligible Mn$^{2+}$ content.

This magnetooptical study was extended to a whole series of samples. As expected, the giant Zeeman shift increases with the Mn$^{3+}$ content
determined from $d$-$d$ absorption. In Fig.~\ref{fig:shift}(b), it is compared as usual to the effective concentration $x_\textrm{eff}=x(1-x)^{12}$
rather than the total concentration $x$ in order to account for the small non linear variation of A Exciton redshift. $x_{eff}$ describes the density
of Mn impurities with no magnetic ion over the 12 nearest-neighbors present in the Ga sublattice.


The redshift of exciton $A$ in $\sigma^-$ polarization unambiguously points to a negative sign of $N_0(\alpha-\beta)$. So far, $N_0 \alpha$ has been
reported to be small ($<0.3$~eV) and almost constant for all DMS, and there are no theoretical hint that the case of GaN-based DMS could be
different. Then, $N_0(\alpha-\beta)$=-1.2~eV means that $\beta$ is positive for GaN:Mn$^{3+}$. If we assume a usual value, $N_0\alpha=0.2\pm0.1$~eV,
we deduce $N_0\beta = +1.4\pm0.3$~eV, which means a strong ferromagnetic interaction between the magnetic ions and the holes. A positive $\beta$ was
rarely observed in DMS, but \cite{Mac93,Herb98} for Cr in II-VI DMS, where Cr has the same $3d^4$ electronic configuration as Mn$^{3+}$ in GaN. This
suggests that a less than a half-filled $d$-shell could be responsible for a positive $\beta$. However this analogy could be misleading, because the
energy position of donor and acceptor levels governs the sign of $\beta$ (Refs.~\onlinecite{Blin01,Bhat94}). As ionization levels are at higher
energy in Mn$^{2+}$ than in Mn$^{3+}$ (see Ref.~\onlinecite{Graf03}), we expect that Mn$^{2+}$ in GaN exhibit also a positive $\beta$, despite its
$3d^5$ electronic configuration. It is interesting to note however that observed magnetooptical giant Zeeman effect is in disagreement with the
negative value of $N_0\beta$ deduced from photoemission experiments\cite{Hwan05} analyzed assuming an incorporation of Mn atoms as Mn$^{2+}$.


To sum up, the giant Zeeman effect of the $A$ and $B$ excitons has
been observed by reflectivity and absorption on a series of
Ga$_{1-x}$Mn$_{x}$N samples. It is well described using an excitonic
model valid for a DMS with the wurtzite structure, and values of the
$s$,$p$-$d$ exchange integrals such that
${N_0(\alpha-\beta)}=-1.2\pm0.2$~eV. Assuming a reasonable value of
$\alpha$ we find a positive sign of $\beta$, which means a
ferromagnetic interaction between Mn$^{3+}$ and holes in GaN.


We thank P. Sati, A. Stepanov for valuable data obtained by EPR technique. We acknowledge W.~Bardyszewski, T. Dietl, and P. Kacman for helpful discussions. The work done in Grenoble was carried out in the CNRS-CEA-UJF Joint Group "Nanophysique et
semiconducteurs". This work was partially supported by Polish Committee for Scientific Research (Grants No. N202 006 31/0153 and N202 123 31/1953)
and by the French Ministry of Foreign Affairs.
%
%


\begin{thebibliography}{99}
%
%
\bibitem{dilGaAs} R. C. Myers \emph{et al.}, Phys. Rev. Lett. {\bf95}, 017204 (2005).
\bibitem{complexGaAs} J. Szczytko \emph{et al.}, Phys. Rev. B {\bf59}, 12935 (1999).
\bibitem{Okab98} J. Okabayashi \emph{et al.}, Phys. Rev. B {\bf 58}, R4211 (1998).
\bibitem{Hwan05} J. I. Hwang \emph{et al.}, Phys. Rev. B {\bf 72}, 085216 (2005).
%
\bibitem{Agga83} R. L. Aggarwal \emph{et al.}, Phys. Rev. B {\bf 28}, 6907 (1983).
\bibitem{Kuro03}S. Kuroda \emph{et al.}, Appl. Phys. Lett. {\bf 83}, 4580 (2003).
\bibitem{Sari06} E. Sarigiannidou \emph{et al.}, Phys. Rev. B {\bf74}, 041306(R) (2006).
\bibitem{Marc06b} S. Marcet \emph{et al.}, Phys. Rev. B {\bf74}, 125201 (2006).
%
\bibitem{Wolo04b} A. Wolos \emph{et al.}, Phys. Rev. B {\bf70}, 245202 (2004).
%
\bibitem{Ando03} K. Ando, Appl. Phys. Lett. {\bf 82}, 100 (2003).
%
%
\bibitem{Titov05} A. Titov \emph{et al.}, Phys. Rev. B {\bf72}, 115209 (2005).
%
\bibitem{Pascal06b} P. Sati and A. Stepanov, private communication.
%
\bibitem{Gil95} B. Gil, O. Briot, and R. L. Aulombard, Phys. Rev. B 52, R17 028, (1995).
\bibitem{Camp97} J. Campo \emph{et al.}, Phys. Rev. B {\bf 56}, R7108 (1997).
\bibitem{Step99} R. St\c{e}pniewski \emph{et al.}, Phys. Rev. B {\bf 60}, 4438 (1999).
%
\bibitem{Lago81} J. Lagois, Phys. Rev. B {\bf 23}, 5511 (1981).
%
\bibitem{Juli98} M. Julier \emph{et al.}, Phys. Rev. B {\bf 57}, R6791 (1998).
\bibitem{Bhat94} A. K. Bhattacharjee, Phys. Rev. B {\bf49}, 13987 (1994).
%
%
\bibitem{Pacu06a} W. Pacuski \emph{et al.}, Phys. Rev. B {\bf 73}, 035214 (2006).
\bibitem{Pacu06c} W. Pacuski \emph{et al.}, Acta Phys. Pol. A, 110, 303 (2006).
%
\bibitem{Mac93} W. Mac \emph{et al.}, Phys. Rev. Lett. {\bf71}, 2327 (1993).
\bibitem{Herb98} M. Herbich \emph{et al.},  Phys.~Rev.~B {\bf 58}, 1912 (1998).
%
\bibitem{Blin01} J. Blinowski and P. Kacman, Acta Phys. Pol. A 100, 343 (2001).
\bibitem{Graf03} T. Graf, S. T. B. Goennenwein, M. S. Brandt, Phys. Status Solidi B, 239, 267 (2003).
\end{thebibliography}
\end{document}